\documentclass[english]{article}
\usepackage[T1]{fontenc}
\usepackage[latin1]{inputenc}
\usepackage{float}
\usepackage{graphicx}
\usepackage{amssymb}

\makeatletter

\usepackage{babel}
\makeatother
\begin{document}

\title{Creation of Entanglement and Implementation of Quantum Logic Gate
Operations Using a Three-Dimensional Photonic Crystal Single-Mode
Cavity}

\date{Durdu Ö. Güney%
\footnote{\noindent Department of Electrical and Computer Engineering, University
of California, San Diego, 9500 Gilman Dr., La Jolla, California, 92093-0409
USA%
}~$^{,}$%
\footnote{\noindent Department of Mathematics, University of California, San
Diego, 9500 Gilman Dr., La Jolla, California, 92093-0112 USA%
} ~and David A. Meyer$^{\dagger}$}

\maketitle
\begin{abstract}
\noindent We solve the Jaynes-Cummings Hamiltonian with time-dependent
coupling parameters under dipole and rotating-wave approximation for
a three-dimensional (3D) photonic crystal (PC) single mode cavity
with a sufficiently high quality (Q) factor. We then exploit the results
to show how to create a maximally entangled state of two atoms, and
how to implement several quantum logic gates: a dual-rail Hadamard
gate, a dual-rail NOT gate, and a SWAP gate. The atoms in all of these
operations are syncronized, which is not the case in previous studies
{[}1,2{]} in PCs. Our method has the potential for extension to $N$-atom
entanglement, universal quantum logic operations, and the implementation
of other useful, cavity QED based quantum information processing tasks.
\end{abstract}

\section{Introduction}

The superiority of quantum computing over classical computation for
problems with solutions based on the quantum Fourier transform, as
well as search and (quantum) simulation problems has attracted increasing
attention over the last decade. Despite promising developments in
theory, however, progress in physical realization of quantum circuits,
algorithms, and communication systems to date has been extremely challenging.
Major model physical systems include photons and nonlinear optical
media, cavity QED devices, ion traps, and nuclear magnetic resonance
(NMR) with molecules, quantum dots, superconducting gates, and spins
in semiconductors {[}3{]}. 

In the last century, control over electrical properties of materials
led to the transistor revolution in electronics. Now, in this century,
PCs inspired great interest for controlling the flow of light. PCs
are periodic arrays of dielectric materials, which would resemble
the atomic structure of semiconductors if the dielectric materials
were scaled down to atomic dimensions. If a defect is introduced by
disturbing the periodicity of the crystal, a localized mode is allowed
in that region. Similar to electronic band structure in semiconductors,
acceptor or donor types of defect states can also be produced in a
PBG, by locally removing or adding extra dielectric material, respectively.
By tuning the parameters of the PC cavity, the frequency or the spatial
profile of this localized state can be tailored on demand. 

Known also as photonic band gap (PBG) materials, PCs seem to be especially
promising as a base medium both for future laser based photonic integrated
circuits and perhaps later on for advanced quantum network technologies
as well {[}4-6{]}. Combining a high Q factor and an extremely small
mode volume successfully in microcavities, PCs have already become
an especially attractive paradigm for quantum information processing
experiments in cavity QED {[}1,2,7{]}. 

Quantum logic gates and quantum entanglement constitute two building
blocks, among others, of more sophisticated quantum circuits and communication
protocols. The latter also allows us to test basic postulates of quantum
mechanics.

In this work we propose an alternative method for creating entanglement
and implementing certain quantum logic gates based on single mode
PC microcavities. Our scheme requires high-quality PCs with Q-factor
of around $10^{8}$. Two-dimensional PC slabs, which have been demonstrated
recently, have a Q-factor of only about $10^{4}$-$10^{5}$. Since
today's planar PC technology does not allow sufficiently high-Q cavities
for our purposes, we must consider three-dimensional PCs despite the
relative difficulty of their fabrication.

\section{Theory}

In this paper we first explore the possibility of mutually entangling
two Rb atoms by exploiting their interaction, mediated by a single
defect mode confined to a three-dimensional PC cavity. We then describe
various logic gates that can be implemented using the same interaction.
We achieve this in three steps: analysis of a single mode cavity (i)
in a generic 3D PC, (ii) in a specific 2D PC and (iii) ultimately
in a 3D version of the 2D PC of (ii). We assume that the defect frequency
is resonant with the Rb atoms. Because the atoms are moving, the time-dependent
Hamiltonian (Jaynes-Cummings Model) for this interaction in the dipole
and rotating wave approximations is

\begin{equation}
H(t)=\frac{\hbar\varpi}{2}\sum_{j}\sigma_{z}^{j}+\hbar\varpi\alpha^{\dagger}\alpha+\hbar\sum_{j}(G_{j}(t)\sigma_{+}^{j}\alpha+h.c.)\end{equation}
 where the summation is over two atoms, $A$ and $B$, $\omega$ is
the resonant frequency, $\sigma_{z}$ is the $z$-component of the
Pauli spin operator and $\sigma_{\pm}$ are atomic raising and lowering
operators. $\alpha$ and $\alpha^{\dagger}$ are photon destruction
and construction operators, respectively. The time-dependent coupling
parameters can be expressed as {[}1{]}

\begin{equation}
G_{j}(t)=\Omega_{0}f_{j}(t)\cos(\zeta_{j})\end{equation}
where $\Omega_{0}$ is the peak atomic Rabi frequency over the defect
mode and $f_{j}(t)$ is the spatial profile of the defect state observed
by the atom $j$ at time $t$. $\zeta_{j}$ is the angle between the
atomic dipole moment vector, $\mathbf{\mu}_{\mathbf{eg}}^{j}$, for
atom $j$ and mode polarization at the atom location.

We ignore the resonant dipole-dipole interaction (RDDI), because the
atoms have their transition frequencies close to the center of a wide
PBG and the distance between them is always sufficiently large that
RDDI effect is not significant {[}2,8{]}.

Initially we prepare one of the atoms, $A$, in the excited state
and the cavity is left in its vacuum state, so

\begin{equation}
|\Psi(0)\rangle=|100\rangle.\end{equation}

The PC should be designed to allow the atoms to go through the defect.
This can be achieved by injecting atoms through the void regions of
a PC with a defect state of acceptor type. Since the spontaneous emission
of a photon from the atoms is suppressed in the periodic region of
the crystal, no significant interaction occurs outside the cavity.
When the atoms enter the cavity, the interaction between the atoms
is enhanced by the single mode cavity. This atom-photon-atom interaction
allows us to design an entanglement process between atoms. 

Given the initial state (3), the state of the system at time $t$
should be in the form

\begin{equation}
|\Psi(t)\rangle=a(t)|100\rangle+b(t)|010\rangle+\gamma(t)|001\rangle\end{equation}
to satisfy the probability and energy conservation. We will show analytically
that the amplitudes at time $t$ can be expressed in terms of the
coupling parameters and the velocities of the atoms.

We can write the Schr\"odinger equation for the time-evolution operator
in the form {[}9{]}

\begin{equation}
i\hbar\frac{\partial}{\partial t}U(t,t_{0})=H(t)U(t,t_{0})\end{equation}

In the basis $\{|100\rangle,|010\rangle,|001\rangle\}$, the matrix
elements of the Hamiltonian (1) are

\begin{equation}
H_{11}=H_{12}=H_{21}=H_{22}=H_{33}=0\end{equation}

\begin{equation}
H_{13}=H_{31}=\hbar G_{A}(t)\end{equation}

\begin{equation}
H_{23}=H_{32}=\hbar G_{B}(t)\end{equation}

Hamiltonian operator, $H(t)$, at different times commute, if $G_{B}(t)$
is a multiple of $G_{A}(t)$ by a factor of $p$. This condition can
be easily satisfied by the appropriate orientation of atomic dielectric
moment of the incoming atoms with respect to electric field in equation
(2). Then the formal solution to (5) becomes

\begin{equation}
U(t,t_{0})=e^{-\frac{i}{\hbar}\int_{t_{0}}^{t}d\tau H(\tau)}=e^{-\frac{i}{\hbar}I},\end{equation}
where $I$ is defined as the integral of the Hamiltonian operator.
By expanding the exponential (9) we obtain

\begin{equation}
U(t,t_{0})=1+(\frac{-i}{\hbar})I+\frac{1}{2!}(\frac{-i}{\hbar})^{2}I^{2}+...+(\frac{1}{n!})(\frac{-i}{\hbar})^{n}I^{n}+...\end{equation}
Multiplying equation (10) by the initial state $|100\rangle$ from
the right and comparing with (4) gives

\begin{equation}
a(t)=1+G_{A}^{2}\sum_{n=1}(-1)^{n}\frac{1}{2n!}(G_{A}^{2}+G_{B}^{2})^{n-1}\end{equation}

\begin{equation}
b(t)=G_{A}G_{B}\sum_{n=1}(-1)^{n}\frac{1}{2n!}(G_{A}^{2}+G_{B}^{2})^{n-1}\end{equation}

\begin{equation}
\gamma(t)=iG_{A}\sum_{n=1}(-1)^{n}\frac{1}{(2n-1)!}(G_{A}^{2}+G_{B}^{2})^{n-1},\end{equation}
where we have defined

\begin{equation}
G_{j}=\int_{t_{0}}^{t}G_{j}(\tau)d\tau.\end{equation}
Recognizing the Taylor series for sine and cosine allows us to rewrite
equations (11)-(13) as

\begin{equation}
a(t)=1+\frac{G_{A}^{2}}{G_{A}^{2}+G_{B}^{2}}[\cos(G_{A}^{2}+G_{B}^{2})^{1/2}-1]\end{equation}

\begin{equation}
b(t)=\frac{G_{A}G_{B}}{G_{A}^{2}+G_{B}^{2}}[\cos(G_{A}^{2}+G_{B}^{2})^{1/2}-1]\end{equation}

\begin{equation}
\gamma(t)=-i\frac{G_{A}}{(G_{A}^{2}+G_{B}^{2})^{1/2}}\sin(G_{A}^{2}+G_{B}^{2})^{1/2}\end{equation}

\begin{figure}[!t]
\includegraphics[%
  bb=15bp 244bp 594bp 841bp,
  scale=0.6]{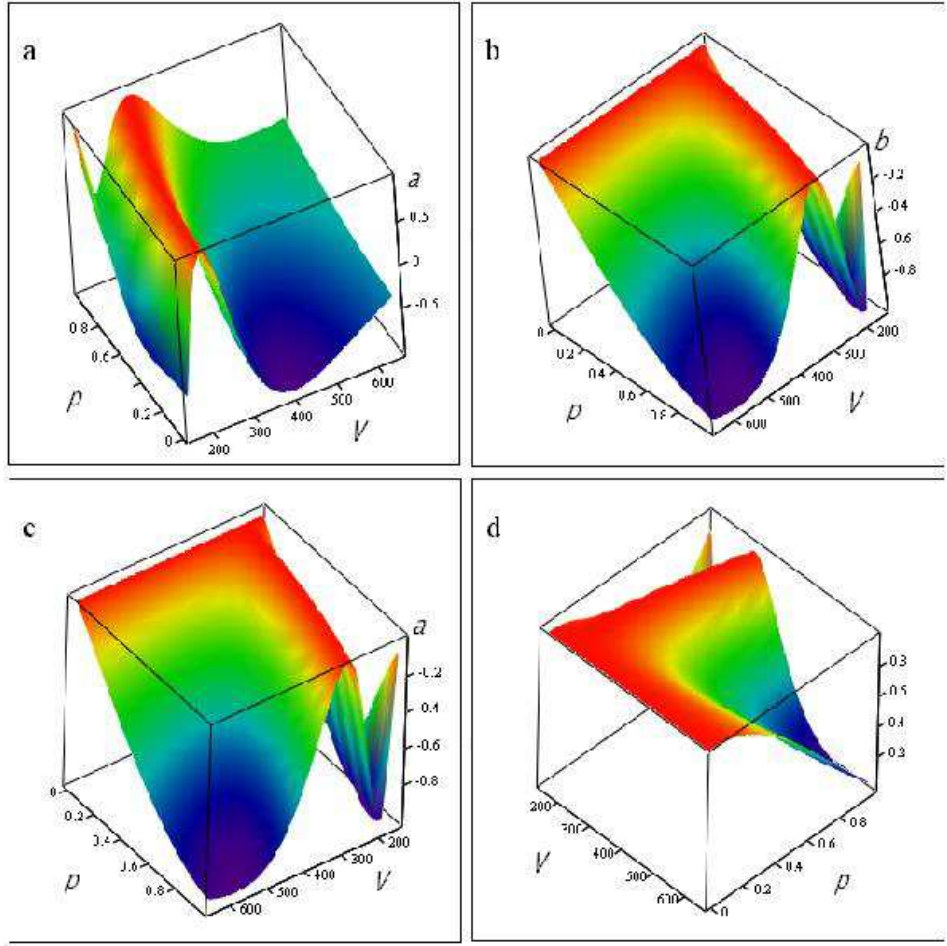}

\caption{Probability amplitudes as a function of the velocity and $p$ (i.e.,
red positive, blue negative). (a) Surface $a(V,p)$ and (b) $b(V,p)$
if the initial state is $|100\rangle$. (c) Surface $a(V,p)$ and
(d) $b(V,p)$ if the system is initially prepared in the $|010\rangle$
state.}
\end{figure}

For a single initially excited atom, $A$, passing across the cavity
we can set $G_{B}=0$ in equations (15)-(17), which gives the same
result with equation (15) of {[}4{]}. The exact solution for the time-independent
problem of $N$ identical two level atoms with a resonant single mode
quantized field given in ref. {[}10{]} could be helpful to generalize
our results to $N$-atom case, see for example ref.\ {[}2{]}.

Absent a rigorous calculation of the defect mode in the three-dimensional
PC, let us just assume a generic spatial profile for the mode, which
oscillates and decays exponentially. Thus, $f_{j}(t)$ in (2) can
be expressed as {[}1{]}

\begin{equation}
f_{j}(t)=e^{-\frac{|V_{j}t-L|}{R_{def}}}\cos[\frac{\pi}{l}(V_{j}t-L)]\end{equation}
where $V_{j}$, $L$, $R_{def}$, $l$ are the velocity of atom $j$,
the total path length of the atoms, defect radius, and the lattice
constant of the PC, respectively.

We choose the velocity of atoms, $V_{j}=V$, such that $150{\rm m/s}<V<650{\rm m/{\rm s}}$,
a typical velocity range appropriate for both experiments and our
calculations. This has two immediate advantages: First, we make Hamiltonians
at different times commute, and thus greatly simplify the analytical
analysis. Second, we syncronize the atoms, providing cyclical readout
that could be synchronized with the cycle time of a quantum computer
{[}24{]}. These conditions in previous studies {[}1,2{]} are not satisfied.

Using equations (15)-(18) we can express the asymptotic probability
amplitudes, $a(t)$ and $b(t)$, as functions of the $V$ and the
$p$. The results for $a(V,p)$ and $b(V,p)$, when the initial state
is $|100\rangle$, are displayed in Figures 1a and 1b, respectively.
If the initial state is $|010\rangle$, corresponding probability
amplitudes are illustrated in Figures 1c and 1d. Note that the surfaces
in Figures 1b and 1c are the same. This is simply due to the symmetry
in equation (16). To compute these surfaces we used the asymptotic
(constant) values of the $G_{j}$. 

\begin{figure}[t]
\includegraphics[%
  bb=15bp 634bp 594bp 841bp,
  scale=0.6]{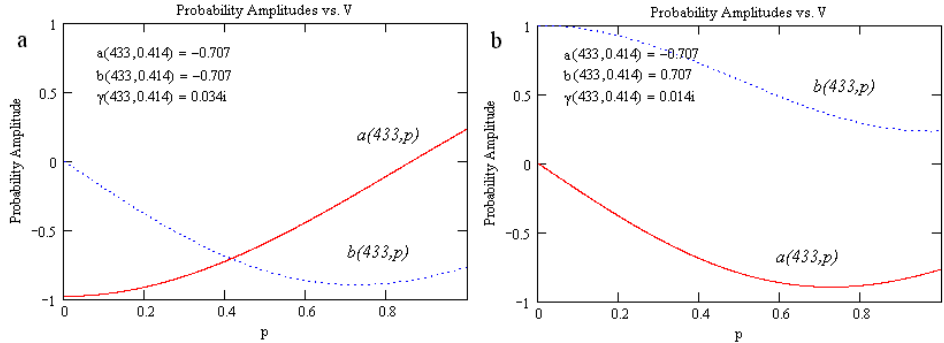}

\caption{Slices from each surface in Fig.\ 1. (a) Probability amplitudes
$a(V,p)$---red---and $b(V,p)$---dashed blue---with $V=433{\rm m}/{\rm s}$.
Entangled state (19) is obtained at $p=0.414$. (b) Probability amplitudes
$a(V,p)$---dashed blue---and $b(V,p)$---red---with the same
velocity, $V=433{\rm m}/{\rm s}$. Entangled state (20) is observed
at the same, $p=0.414$, value.}
\end{figure}

\begin{figure}[t]
\includegraphics[%
  bb=15bp 394bp 594bp 841bp,
  scale=0.6]{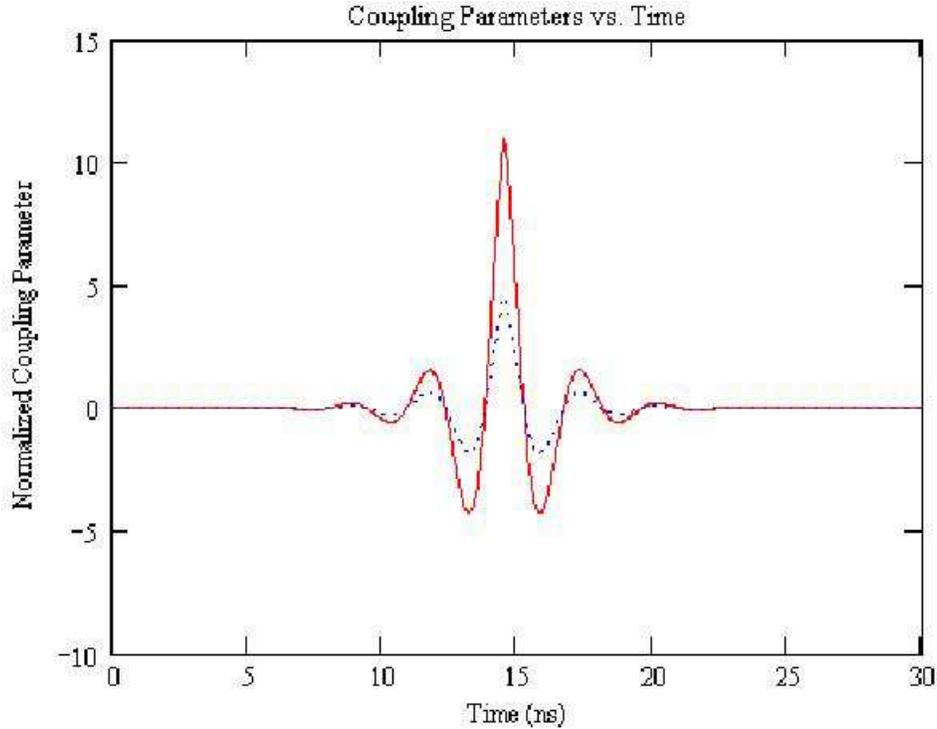}

\caption{Coupling parameters in the reference frame of moving atoms with velocities
$V=433{\rm m}/{\rm s}$ at $p=0.414$. $\Omega_{0}=11x10^{9}{\rm Hz}$,
$\omega=2.4\times10^{15}{\rm Hz}$, $l=1.6\frac{\pi c}{\omega}$,
$L=10l$, $R_{def}=l$ {[}1{]}. Atom $A$ experiences the coupling
parameter shown with the red solid curve and atom $B$ the one shown
with dashed blue. }
\end{figure}

\begin{figure}[t]
\includegraphics[%
  bb=15bp 624bp 594bp 841bp,
  scale=0.6]{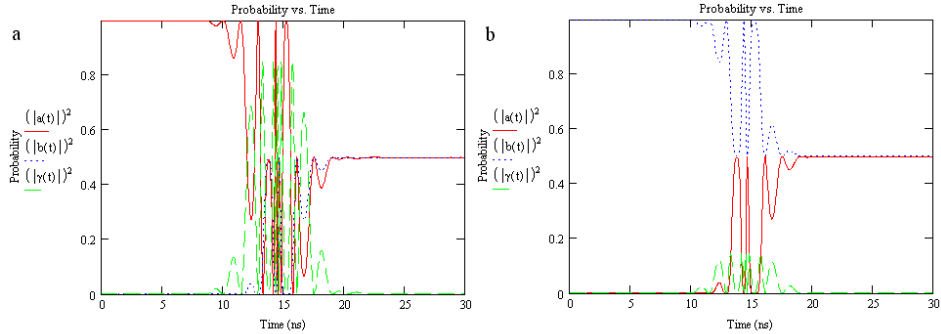}

\caption{Time evolution of the probabilities that shows entanglement when
(a) the initial state is $|100\rangle$ and (b) $|010\rangle$.}
\end{figure}

\begin{figure}
\includegraphics[%
  bb=15bp 634bp 594bp 841bp,
  scale=0.6]{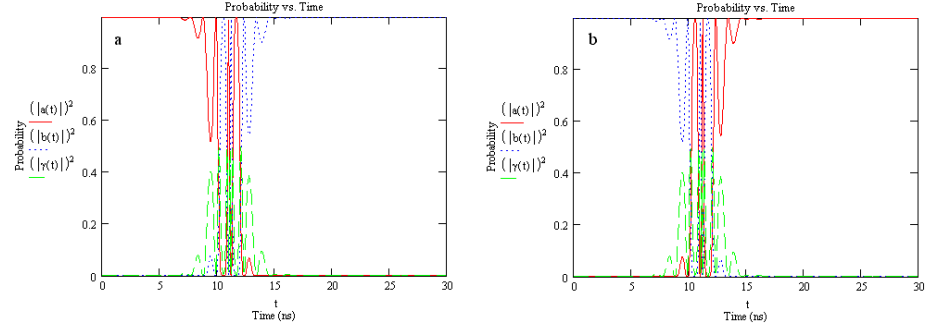}

\caption{Time evolution of the probabilities that leads to a dual-rail NOT
(Pauli $\sigma_x$) logic operation when (a) the initial state is
$|100\rangle$ and (b) $|010\rangle$. }
\end{figure}

Fig.\ 2a illustrates a slice from the surface in Fig.\ 1a where
the velocity of the atoms are, $V=433{\rm m}/{\rm s}$. Note that
we obtain the maximally entangled state,

\begin{equation}
|\Psi_{10}\rangle\cong\frac{|10\rangle+|01\rangle}{\sqrt{2}},\end{equation}
up to an overall phase, $-1$, when the velocities of the atoms, $V=433{\rm m}/{\rm s}$
and the initial state is $|100\rangle$. Similarly if we keep the
velocities of the atoms fixed but set the atom $B$ to be excited
initially (i.e., $|010\rangle$), we obtain the slice in Fig.\ 2b
and thus the maximally entangled state,

\begin{equation}
|\Psi_{01}\rangle\cong\frac{|10\rangle-|01\rangle}{\sqrt{2}},\end{equation}

up to the same overall phase factor as in equation (19). (In equations
(19) and (20), and in the following, we omit the cavity state in the
kets, since it factors out and does not contribute to logic operations
with which we are concerned.)

The calculated coupling parameters in the reference frame of the atoms
as a function of time, $G_{j}(t)$, with the specific values of $\Omega_{0}$,
$L$, $R_{def}$, and $l$ are shown in Fig.\ 3. The solid red curve
and blue dashed curve correspond to atom $A$ and $B$, respectively.
The total interaction time is less than $20{\rm ns}$.

Probabilities $|a(t)|^{2}$, $|b(t)|^{2}$, and $|\gamma(t)|^{2}$
from equations (15)-(17) are graphed in Fig.\ 4. Fig.\ 4a shows
the evolution of the probabilities when the initial state is $|100\rangle$.
Note that the cavity is disentangled from the atoms and we end up
with the final state (19). On the other hand if the initial state
is $|010\rangle$, the time evolution of the probabilities is illustrated
in Fig.\ 4b. Note that the final state in this setting becomes (20),
since the cavity is again disentangled. From (19) and (20) it is clear
that the quantum system we have described not only entangles the atoms
but {\em also operates as a dual-rail Hadamard gate {[}3{]} up to
an overall phase.} 

Using the same quantum system one can also build a dual-rail NOT gate
under certain conditions. If we set $V_{A}=V_{B}=565{\rm m}/{\rm s}$,
for example, we obtain the following logical transformations, up to
an unimportant global phase factor, which defines a dual-rail NOT
gate (see fig.\ 5):

\begin{equation}
|10\rangle\mapsto|01\rangle\end{equation}

\begin{equation}
|01\rangle\mapsto|10\rangle\end{equation}

Furthermore using the Hamiltonian (1) it can be shown that $|00\rangle$
initial state only gain a deterministic phase factor of $-1$ in interaction
picture. Once the conditions (21) and (22) are satisfied $|11\rangle$
initial state is transformed into itself up to the same global phase
with the states in (21) and (22). Thus, including these as possible
initial states our dual-rail NOT gate also operates as a SWAP gate
up to a relative phase of the $|00\rangle$ state. During our analysis
we have also observed that a dual-rail $Z$ gate is also possible
for some specifications.

\section{Two-dimensional Photonic Crystal}

In the preceding analysis we have assumed the generic form (18) for
the spatial profile of the defect mode in order to demonstrate that,
in principle, PC microcavities can be used as entanglers, and more
specifically, as certain logic gates. In the following we apply these
ideas to real two- and three-dimensional photonic crystal microcavity
designs, to show that implementations of these quantum devices are
indeed possible in these photonic systems. As the authors of {[}2{]}
observe, however, ``a rigorous calculation of the electromagnetic
field in the presence of a defect in a 3D photonic crystal can be
a difficult task''. In the following we address this task systematically.

First we consider a 2D photonic crystal design with a triangular lattice
of dielectric rods with dielectric constant of $12$ (i.e., silicon)
in Fig.\ 6. The radius of the rods is $0.175l$, where $l$ is the
lattice constant. The symmetry is broken in the center by introducing
a defect with reduced rod-radius of $0.071l$ to form the microcavity
of our quantum system. We assume that the atoms $A$ and $B$ travel
along the dashed green lines shown in Fig.\ 6a, although any two
of the obvious paths shown by dashed lines (or any void regions with
line of sight of the cavity) work as well. 

\begin{figure}
\includegraphics[%
  bb=15bp 544bp 594bp 841bp,
  scale=0.6]{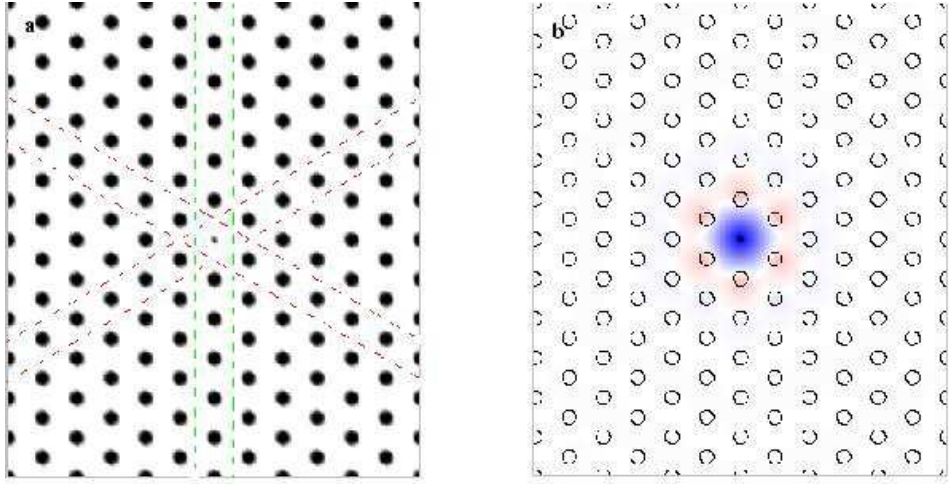}

\caption{(a) A single-mode microcavity in a 2D photonic crystal with a triangular
lattice (see the text for details). (b) Corresponding electric field
spatial profile for the transverse-magnetic (TM) mode allowed in the
cavity.}
\end{figure}

In order to find the probability amplitudes for the atoms as a function
of time while travelling through the crystal, we need to calculate
the coupling parameters in equation (14) and substitute them in equations
(15)-(17). Once we obtain the probability amplitudes, we can demonstrate
entanglement creation and construct logic gates as before. We also
note that the quality factor of the cavity should be high enough that
the interaction time (i.e., the logic operation time) is much less
than the photon lifetime in the cavity. Below we show that typical
logic operations take $50{\rm \mu s}$ for the Hadamard gate and $30{\rm \mu s}$
for the NOT or the SWAP gate. Thus a quality factor of $10^{8}$ should
be sufficient for reliable gate operations. The quality factor of
the cavity could be increased exponentially with additional period
of rods in Fig.\ 6a {[}11{]}. We observe, however, that the spatial
profile of the mode in Fig.\ 6b does not change significantly (and
hence degrade the gate) after a certain number of periods. Thus, the
exact number of periods required for a quality factor of $10^{8}$
is not essential to demonstrate our main goal in this paper, namely
that such logic operations and entanglement creation are indeed possible
in these photonic crystal structures. 

In our generic profile (18), the coupling parameter is assumed to
be real. However, for generality in real photonic crystals, we must
allow it to be a complex parameter. Thus, the interaction part of
the Hamiltonian (1) for a single-atom cavity interaction can be written
as {[}12-15{]}

\begin{equation}
H_{I}=\hbar|g(\mathbf{r})|[\alpha^{\dagger}\sigma_{-}+\alpha\sigma_{+}]\end{equation}

by incorporating complex coupling parameter $g(\mathbf{r})$ into
(1). 

In a photonic crystal we can express the atom-field coupling parameter
{[}12, 16{]} at the position of atom $j$, as

\begin{equation}
g(\mathbf{r}_{j})=g_{0}\Psi(\mathbf{r}_{j})\cos(\zeta_{j})\end{equation}
where $g_{0}$ and $\Psi(\mathbf{r}_{j})$ are defined as

\begin{equation}
g_{0}\equiv\frac{\mu_{eg}}{\hbar}\sqrt{\frac{\hbar\omega}{2\varepsilon_{0}\varepsilon_{m}V_{mode}}}\end{equation}

\begin{equation}
\Psi(\mathbf{r}_{j})\equiv E(\mathbf{r}_{j})/|E(\mathbf{r}_{m})|\end{equation}

$\mathbf{r}_{m}$ denotes the position in the dielectric where $\varepsilon(\mathbf{r})|E(\mathbf{r})|^{2}$
is maximum and $\varepsilon_{m}$ is defined as the dielectric constant
at that point. The cavity mode volume, $V_{mode}$, is given as

\begin{equation}
V_{mode}=\frac{\int\int\int\varepsilon(\mathbf{r})|E(\mathbf{r})|^{2}d\mathbf{r}}{\varepsilon_{m}|E(\mathbf{r}_{m})|^{2}}\end{equation}

\begin{figure}[t]
\includegraphics[%
  bb=15bp 514bp 594bp 841bp,
  scale=0.6]{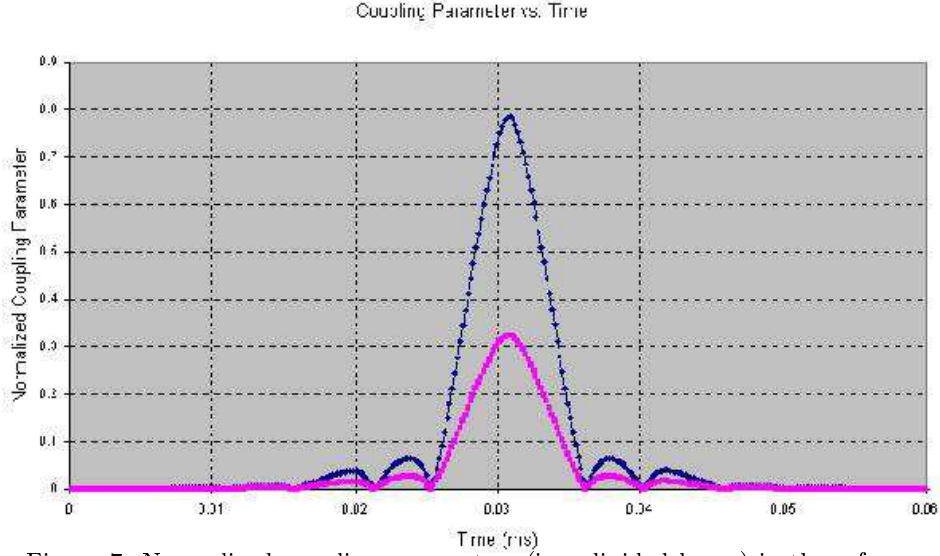}

\caption{Normalized coupling parameters (i.e., divided by $g_{0}$) in the
reference frame of moving atoms with velocities $V=374{\rm m}/{\rm s}$
at $p=0.414$, where $g_{0}$ is found to be $2.765{\rm MHz}$. Blue
and red curves correspond to normalized coupling strengths for atoms
$A$ and $B$, respectively.}
\end{figure}

\begin{figure}[t]
\includegraphics[%
  bb=15bp 684bp 594bp 841bp,
  scale=0.6]{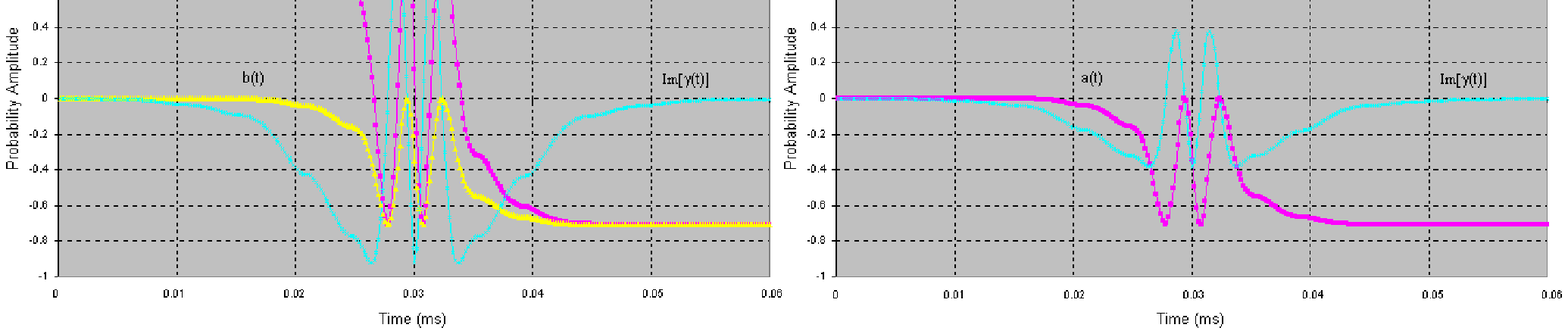}

\caption{Probability amplitudes for entangled atoms created by a dual-rail
Hadamard operation in the 2D photonic crystal (see Fig.\ 6) when
(a) atom $A$ is initially in the excited state and (b) when atom
$B$ is initially in the excited state. $a(t)$, $b(t)$ and $\gamma(t)$
are probability amplitudes for the states $|100\rangle$, $|010\rangle$
and $|001\rangle$, respectively. }
\end{figure}

\begin{figure}
\includegraphics[%
  bb=15bp 684bp 594bp 841bp,
  scale=0.6]{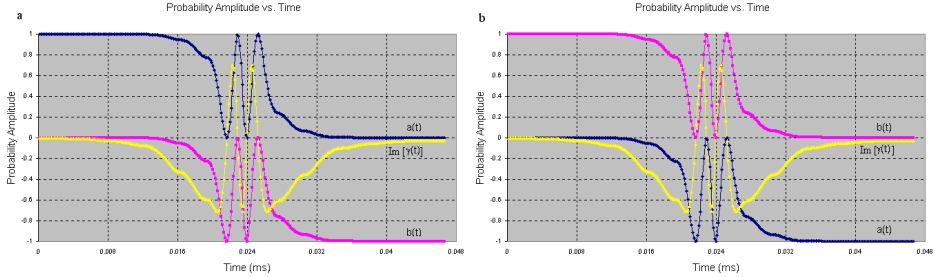}

\caption{Probability amplitudes for atom $A$ and $B$ under the dual-rail
NOT operation in the 2D photonic crystal when initially (a) atom $A$
is excited only and (b) atom $B$ is excited only. }
\end{figure}

Using the block iterative plane-wave expansion method {[}17{]} we
found the normalized frequency of the cavity mode shown in Fig.\
6b to be $0.3733c/l$. By setting $l=2.202{\rm mm}$, we tune the
resonant wavelength to $5.9{\rm mm}$. At this wavelength, $\mu_{eg}$
for the Rb atom is $2\times10^{-26}{\rm C}\cdot{\rm m}$ {[}1{]}.
Since we observe that the energy density is concentrated in the center
of the cavity, $\varepsilon_{m}$ in equations (25) and (27) becomes
$12$. On the other hand, $\cos(\zeta_{j})$ in equation (24) can
be safely assumed constant for each atom, because our cavity mode
is transverse-magnetic (TM) mode and its electric field polarization
direction can always be assumed to have a constant angle with the
atomic dipole moment vector, $\mathbf{\mu_{eg}}^{j}$. We determined
the coupling parameters as functions of time in the reference frame
of moving atoms, as shown in Fig.\ 7. The tails of the coupling parameter
functions do not contribute significantly to results due to exponential
decay away from the cavity. 

With these settings Fig.\ 8 demonstrates an entangler which operates
as a dual-rail Hadamard gate. If the initial state is $|10\rangle$
(i.e., atom $A$ is excited), we end up with state $|\Psi_{10}\rangle$
{[}see equation (19) and Fig.\ 8a{]}. If atom $B$ is initially excited
(i.e., the initial state is $|01\rangle$), however, we obtain the
state $|\Psi_{01}\rangle$, up to an unimportant global phase factor
of $-1$ {[}see equation (20) and Fig.\ 8b{]} as the output of the
logic gate. 

In order to get the system to act as a dual-rail NOT gate, we simply
set the velocities of atoms $V_{A}=V_{B}=490{\rm m}/{\rm s}$. The
evolution of the probability amplitudes for the atoms is shown in
Fig.\ 9. When the excitation is initially at atom $A$ (or at atom
$B$), it is transferred to atom $B$ (or atom $A$). Note that we
also get an unimportant phase factor of $-1$ in the output. Furthermore,
as explained above, the input states $|00\rangle$ (or $|11\rangle$)
are not transformed into different states, but only $|00\rangle$
gain a different phase factor of $+1$. Once this phase problem with
the $|00\rangle$ state is solved, the system could also be exploited
as a SWAP gate.

\section{Three-dimensional Photonic Crystal}

Although the implementation of logic gates in 2D photonic crystals
looks promising, in reality we need 3D devices. The 2D analysis is
useful, however, for reducing the substantial computation required
for analyzing more realistic 3D structures, because there is great
similarity between the modes allowed in these 2D crystals and in their
carefully chosen 3D counterparts (see Fig.\ 10), which we describe
next. 

To demonstrate logic operations in 3D photonic crystals, we designed
the structure {[}18-19{]} shown in Fig.\ 10a. The 3D photonic crystal
we have chosen has various advantages over others {[}20-22{]}: emulation
of 2D properties in 3D {[}19{]}, polarization of the modes, and simplified
design and simulation. It consists of alternating layers of a triangular
lattice of air holes and a triangular lattice of dielectric rods,
where the centers of the holes are stacked along the {[}111{]} direction
of the face-centered cubic (fcc) lattice. The parameters for the crystal
are given in the caption of Fig.\ 10a. We calculated that the structure
exhibits a 3D band gap of over $20\%$, across the frequency range
of $0.507c/l$ - $0.623c/l$.

As in the 2D geometry, we introduce a defect by reducing the radius
of a rod inside the crystal as shown in Fig.\ 10b. Using the supercell
method we computed that this cavity supports only a single mode with
a frequency of $0.539c/l$. Setting $l=3.18{\rm mm}$ tunes the cavity
mode to the atomic transition wavelength of $5.9{\rm mm}$. Thus we
can design a 3D single-mode cavity for our system to operate as the
desired quantum logic gates. 

The spatial profile for the electric field of the mode is shown in
Figs.\ 10c and 10d. Note that it has similar profile with its 2D
counterpart in Fig.\ 6b, where the energy density is also maximized
in the center of the cavity. It is this similarity which simplifies
our design and analysis for the 3D case. 

We can quantify {[}19{]} the TM-polarization of the mode in a plane
as:

\begin{equation}
P\equiv\frac{\int d^{2}\mathbf{r}|E_{z}(\omega;\mathbf{r})|^{2}}{\int d^{2}\mathbf{r}|{\rm E}(\omega;\mathbf{r})|^{2}}.\end{equation}

We compute that $P$ is almost $0.99$ in the defect plane shown in
Fig.\ 10b for the spatial profile exhibited in Figs.\ 10c and 10d.
Then it is safe to assume TM polarized mode in the system Hamiltonian,
because this doesn't affect the probability amplitudes significantly
in the microwave regime, for the parameters we have chosen. 

\begin{figure}[H]
\includegraphics[%
  bb=10bp 374bp 594bp 841bp,
  scale=0.6]{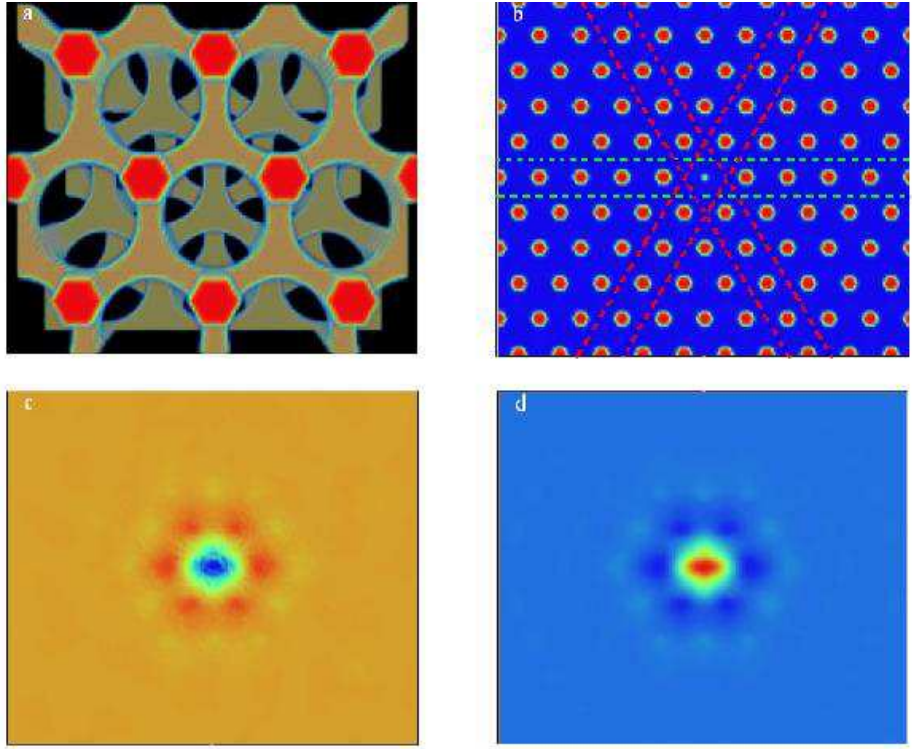}

\caption{(a) Top view of the 3D photonic crystal with fcc lattice. It consists
of alternating layers of a triangular lattice of air holes and a triangular
lattice of dielectric rods (for details of the structure, see refs
{[}18-19{]}). The nearest-neighbor spacing within either a hole or
rod layer is $\frac{1}{\sqrt{2}}l$, where $l$ is the fcc lattice
constant. Hole and rod radii are $0.293l$ and $0.124l$, respectively.
The thicknesses of a hole layer and a rod layer are taken to be $0.225l$
and $0.354l$, respectively. Silicon is assumed as the high-index
material of dielectric constant $12$. (b) Horizontal cross-section
of the crystal. Dashed lines show the obvious paths atoms can travel.
In our simulations we assumed the path shown by the dashed yellow
line. A defect is introduced by reducing the radius of the middle
rod down to $0.050l$ to hold a single mode in the cavity. (c) The
real part and (d) imaginary part of the electric-field of the TM mode
allowed in the cavity at a particular instant in time with the frequency
of $0.539c/l$. The imaginary part is half a period later.}
\end{figure}

The coupling parameters as function of time in the reference frame
of atoms are shown in Fig.\ 11, where the velocities of atoms are
set to be $V=353{\rm m}/{\rm s}$ at $p=0.414$. 

In Fig.\ 12 we demonstrate the 3D version of our 2D dual-rail Hadamard
gate, which also acts as an atomic entangler. Similarly to the 2D
case, if the excitation is on atom $A$, the resulting state is $|\Psi_{10}\rangle$
{[}see equation (19){]} as shown in Fig.\ 12a, while it is on atom
$B$ the output state is $|\Psi_{01}\rangle$ {[}see equation (20){]}
as displayed in Fig.\ 12b, up to an unimportant global phase of $-1$.
Interaction between atom $A$ and $B$ is mediated by the photonic
qubit when correct parameters are set. 

We set the velocity of both atoms to $V_{A}=V_{B}=459{\rm m}/{\rm s}$
to obtain a dual-rail NOT gate in the 3D photonic crystal, up to an
unimportant global phase factor, $-1$. The probability amplitude
evolution of the atoms is displayed in Fig.\ 13. 

\begin{figure}
\includegraphics[%
  bb=15bp 514bp 594bp 841bp,
  scale=0.6]{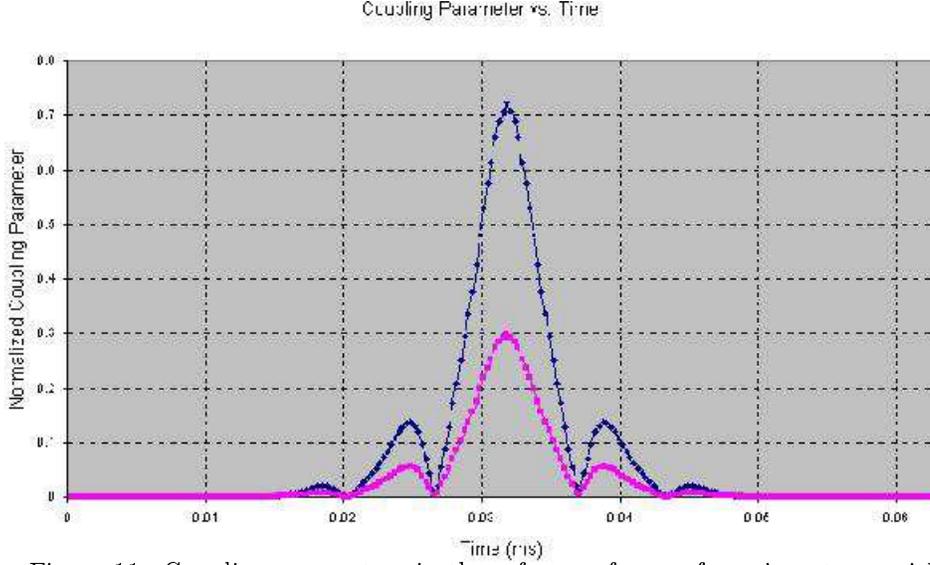}

\caption{Coupling parameters in the reference frame of moving atoms with velocities
$V=353{\rm m}/{\rm s}$ at $p=0.414$, where $g_{0}$ is found to
be \textbf{$2.899{\rm MHz}$}. Blue and red curves correspond to normalized
coupling strengths for atoms $A$ and $B$, respectively.}
\end{figure}

\begin{figure}[t]
\includegraphics[%
  bb=15bp 674bp 594bp 841bp,
  scale=0.6]{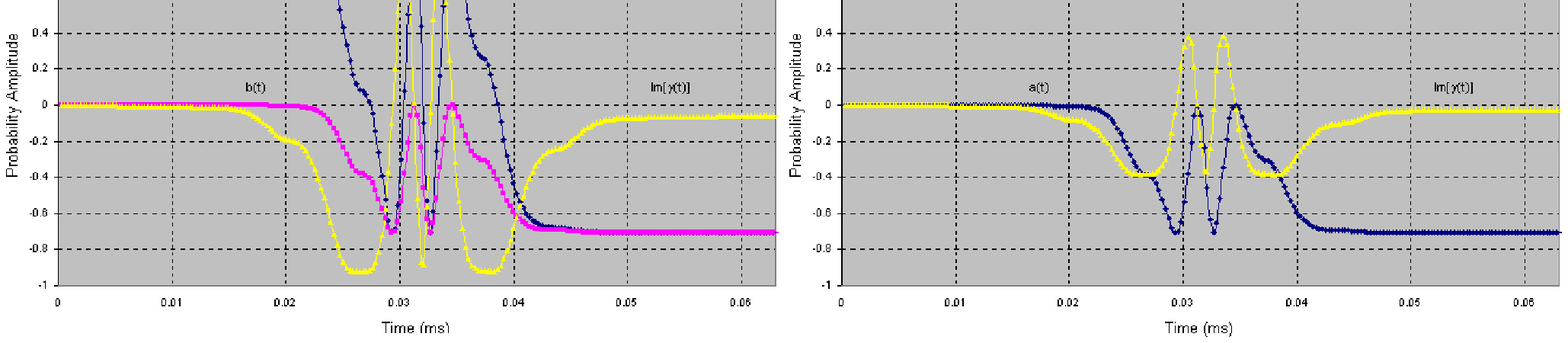}

\caption{Probability amplitudes for entangled atoms under dual-rail Hadamard
operation in the 3D photonic crystal (see Fig.\ 10) when (a) atom
$A$ is initially in the excited state and (b) when atom $B$ is initially
in the excited state. $a(t)$, $b(t)$, and $\gamma(t)$ are probability
amplitudes for the states $|100\rangle$, $|010\rangle$ and $|001\rangle$,
respectively.}
\end{figure}

\begin{figure}
\includegraphics[%
  bb=15bp 674bp 594bp 841bp,
  scale=0.6]{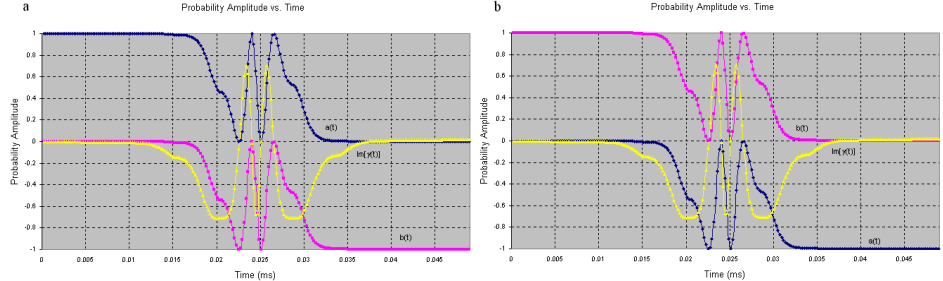}

\caption{Probability amplitudes for atom $A$ and $B$ under dual-rail NOT
operation in the 3D photonic crystal when initially (a) atom $A$
is excited only and (b) only atom $B$ is excited.}
\end{figure}

The excitation of the excited atom is transferred to the ground state
atom with the help of the photonic qubit allowed in the designed cavity.
Note that because of the symmetry of the problem, the $b(t)$s in
Figs.\ 5a, 9a and 13a are indeed approximately the same as the $a(t)$s
of Figs.\ 5b, 9b and 13b, respectively. Furthermore, for the same
reason as in the 2D case, analyzed above, our 3D dual-rail NOT gate
also operates as a SWAP gate up to some deterministic phase. That
is,

\begin{equation}
|00\rangle\mapsto-|00\rangle\end{equation}

\begin{equation}
|01\rangle\mapsto|10\rangle\end{equation}

\begin{equation}
|10\rangle\mapsto|01\rangle\end{equation}

\begin{equation}
|11\rangle\mapsto|11\rangle\end{equation}

Note also that the velocities of the atoms and $g_{0}$ for the 2D
logic gates agree more than $90\%$ with those in 3D. This makes sense
due to the fact that the localized modes allowed in 2D and 3D cavities,
respectively, which we considered above, have more than $90\%$ overlap
in their spatial profiles {[}19{]}. Thus our results are also consistent
with ref.\ {[}19{]} and this is another reason, why we first investigated
2D case.

\section{Conclusions}

To summarize, photonic band gap materials could be especially promising
as robust quantum circuit boards for the delicate next generation
quantum computing and networking technologies. The high quality factor
and extremely low mode volume achieved successfully in microcavities
have already made photonic crystals an especially attractive paradigm
for quantum information processing experiments in cavity QED {[}1,2,7{]}.
In our paper we have extended this paradigm by solving analytically
the Jaynes-Cummings Hamiltonian under dipole and rotating wave approximation
for two syncronized two-level atoms moving in a photonic crystal and
by applying it to produce the two maximally entangled states in equations
(19) and (20). We have also demonstrated the design of quantum logic
gates, including dual-rail Hadamard and NOT gates, and SWAP gate operations.
Our proposed system is quite tolerant to calculation and/or fabrication
errors. Furthermore, most errors left beyond the design can also be
easily circumvented by simply adjusting the velocity or the angle
between the atomic moment vector of atoms and the mode polarization
in experiments. Our technique could not only be generalized to $N$-atom
entanglement {[}2{]} but also has potential for universal quantum
logic gates, atom-photon entanglement processes, as well as the implementation
of various, useful cavity QED based quantum information processing
tasks. We should also mention the methodological result that thanks
to the emulation of 2D photonic crystal cavity modes in 3D photonic
crystals {[}19{]}, one can design the more sophisticated circuit first
in 2D to reduce difficulty of the 3D computations where typically
much more computational power is needed.

\end{document}